\documentclass[11pt,a4paper,english]{article}
\usepackage{amsmath,latexsym,amssymb,amsthm,color}
\usepackage{fullpage}
\usepackage{ifthen,graphics,epsfig}
\usepackage{times}
\usepackage{graphicx,url}
\usepackage{babel}

\newcommand{\REG}{\mathit{REG}}
\newcommand{\Xomit}[1]{}

\newtheorem{theorem}{Theorem}

\newcommand{\toto}{xxx}
\newenvironment{proofT}{\noindent{\bf
Proof }} {\hspace*{\fill}$\Box_{Theorem~\ref{\toto}}$\par\vspace{3mm}}

\newcounter{linecounter}
\newcommand{\linenumbering}{\ifthenelse{\value{linecounter}<10}{(0\arabic{linecounter})}{(\arabic{linecounter})}}

\renewcommand{\thelinecounter}{\ifnum \value{linecounter} > 9\else 0\fi \arabic{linecounter}}

\title{\bf A Necessary Condition for Byzantine  $k$-Set Agreement}

\author{Zohir Bouzid$^{\circ}$, ~~
        Damien Imbs$^{\dag}$, ~~ 
        Michel Raynal$^{\circ,\star}$
~\\~\\
$^{\circ}$  IRISA, Universit\'e de Rennes,  France \\
$^{\dag}$  Dpt of Mathematics, University of Bremen, Germany\\
$^{\star}$  Institut Universitaire de France\\
{\small{\tt  zohir.bouzid@gmail.com ~ imbs@uni-bremen.de ~ raynal@irisa.fr}}\\~\\
Tech Report 2029, IRISA, University of Rennes, December 2015
}

\date{}

\begin{document}

\maketitle
\vspace{-0.5cm}
\begin{abstract}
This short paper presents a necessary condition for Byzantine $k$-set agreement 
in (synchronous or asynchronous) message-passing systems and
asynchronous shared memory systems where the processes communicate through 
atomic single-writer multi-reader registers. 
It gives a proof,  which is particularly  simple, that $k$-set agreement
cannot be solved $t$-resiliently in an $n$-process system when 
$n \leq 2t + \frac{t}{k}$.
This bound  is tight for the case $k=1$ (Byzantine consensus) in 
synchronous message-passing systems.   
\end{abstract}



\section{Computation  Models}

\paragraph{Process failure model}
The system is made up of $n$ sequential processes $p_1$, $_2$, ..., $p_n$. 
A process that executes without deviating  from its intended behavior
(as defined by its algorithm) is said to be {\it correct} or {\it non-faulty}. 
A process that  deviates from its intended behavior is {\it faulty}.
If it can deviate arbitrarily, it is {\it Byzantine}~\cite{LSP82}. 
This is the fault model considered in this paper. 
The parameter $t$ denotes the maximum number of processes 
which may be Byzantine in the corresponding computation model.

The bad behavior of a Byzantine process can be intentional or not. 
As examples of Byzantine behaviors, a  process may crash, 
fail to communicate, communicate fake  values, communicate correctly 
with some processes and incorrectly with others, etc. Several
Byzantine processes can also collude to ``pollute'' the computation.

\paragraph{Message-passing communication models}
We consider here two types of communication media and three computation models. 
The first  communication medium is the classical message-passing model. 
Each pair of processes is connected by a bi-directional channel. 
To simplify the presentation, it is assumed that each process has a
channel from itself to itself. Moreover, the channels are reliable 
(there is neither loss, corruption, duplication, nor creation of messages). 
The message-passing medium gives rise to two computation models. 
\begin{itemize}
\vspace{-0.2cm}
\item  The  asynchronous model, denoted ${\cal BAMP}_{n,t}$,   
considers  $n$ asynchronous processes, among which up to $t$ may be
Byzantine, and  asynchronous channels. ``Asynchronous'' means that 
(a) each process proceeds according to its own speed which may vary with time and 
always remains unknown to the other processes, and (b) the transit time of 
each message is  finite but unbounded.  
\vspace{-0.2cm}
\item  The synchronous model, denoted ${\cal BSMP}_{n,t}$  
considers  $n$  processes, among which up to $t$ may be Byzantine,
which execute a sequence of rounds in a lock-step manner. 
In every round a process first sends messages, then receives messages, 
and finally executes a local  computation. The important property is that 
a message sent in a round is received in the very same round. 
\end{itemize}

\paragraph{Read/write register communication model}
The second communication medium  is the shared memory model where the processes
communicate through single-writer multi-reader (SWMR) atomic registers. 
The processes are asynchronous (as in the corresponding message-passing model). 
An SWMR register is a read/write register that can be written by a single 
predefined process, and read by any process. 
It follows that no Byzantine process can write into a register whose  writer 
is a correct process. 
Let us observe that the use of SWMR atomic registers is natural 
in the presence of Byzantine processes: using multi-writer multi-reader
registers would allow  Byzantine processes to corrupt the whole memory, 
so that no ``useful'' computation could be done. 
The corresponding model is denoted ${\cal BARW}_{n,t}$.

\section{The $k$-Set Agreement Problem}

\paragraph{Definition}
The $k$-set agreement problem is a generalization of the consensus problem, 
which corresponds to the case $k=1$.   
It was introduced by S. Chaudhuri~\cite{C93} in  the context of the process 
crash failure model. A crash is an unexpected stop without recovery.
The aim was to investigate the relation between the maximal number of faulty 
processes ($t$) and the the minimal number of allowed decision values ($k$).

The problem consists in providing the processes with an operation
${\sf propose}_k()$, which returns a value to the invoking process. 
According to the usual terminology, when $p_i$ invokes ${\sf propose}_k(v_i)$,
we say ``$p_i$ proposes  value $v_i$''. If the invocation returns $v$, we 
say ``$p_i$ decides $v$''. 

The $k$-set agreement problem is defined by the following properties
(which means that any algorithm solving the problem must satisfy them).  
\begin{itemize}
\vspace{-0.1cm}
\item Termination. 
The invocation of ${\sf propose}_k()$ by a  correct process terminates. 
\vspace{-0.2cm}
\item Agreement. 
At most $k$ different values are decided by correct processes. 
\vspace{-0.2cm}
\item Validity. 
If all correct  processes propose the same value, no other value 
can be decided by a correct process. 
\end{itemize}

\paragraph{On the validity property}
The validity property relates the outputs (values decided by the correct 
processes) to the inputs (values proposed by the correct processes). 
Let us notice that the previous validity property is particularly weak. 
As soon as two correct processes propose different values, any set of at 
most $k$ (possibly arbitrary) values can be collectively decided by 
the correct processes~\cite{PMR01}.

Stronger validity properties could be considered, such as: 
a value decided by a correct  was proposed by a correct process.
The interest of the weaker validity property lies in the fact that
it enlarges the scope of our necessary condition on $t$. 
To be implemented, any stronger validity property 
requires a constraint on $t$ as strong or even stronger  than our 
condition~\cite{CFV06,HKR14,L96,MMR16,MR10}.

\vspace{-0.2cm}
\section{A Necessary Condition for $k$-Set Agreement in 
 ${\cal BAMP}_{n,t}$ and  ${\cal BSMP}_{n,t}$}
\vspace{-0.1cm}
\begin{theorem}
\label{theo-necessity}
There is no algorithm that solves $k$-set agreement
in  ${\cal BAMP}_{n,t}$ or  ${\cal BSMP}_{n,t}$ when $n\leq 2t+\frac{t}{k}$. 
\end{theorem}

\begin{proofT}
The proof is made up of two parts. 

\noindent
Part 1 on the proof. \\
Let $\Sigma$ be an $n$-process system  such that $n\leq 2t+\frac{t}{k}$, 
and $C$ ($F$) be its set of correct (faulty) processes.  
Assuming  $|C|\leq t+\frac{t}{k}$ and $|F|=t$, 
Let us partition the set $C$ composed of all correct processes into $(k+1)$ 
subsets  $S_1$, ..., $S_{k+1}$,  such that any of these subsets contains 
$\lfloor\frac{n-t}{k+1}\rfloor$ or $\lceil\frac{n-t}{k+1}\rceil$ processes 
(hence, $\forall~i,j \in [1..(k+1)]:~|S_i|-|S_j|\leq 1$).  
This system is represented in the left part of 
Figure~\ref{fig:system-sigma1-gamma}, where 
a segment connecting two sets means that each process of a set is connected to
each process of the other set (remember that the message-passing 
communication graph is complete). Let $\overline{S_i}=C\setminus S_i$. 

\noindent
Claim: $|\overline{S_i}|\leq t$. \\
Proof of the claim. 
Let us assume by contradiction that $|\overline{S_i}|> t$.
As $S_i$ and $\overline{S_i}$ define a partition of $C$, we have 
$|S_i|+|\overline{S_i}|=|C|\leq t+\frac{t}{k}$. 
As $|\overline{S_i}|> t$, it follows that $|S_i|<\frac{t}{k}$. 
Moreover, as $\overline{S_i}$ contains $k$ sets (all subsets $S_x$ of $C$ 
except $S_i$), and their cardinality differ at most by $1$, 
there is necessarily a subset 
$S_j\in \overline{S_i}$ such that $|S_j| > \frac{t}{k}$. 
For the same cardinality reason,  it follows from 
$|S_j|> \frac{t}{k}$   that $|S_i|\geq \frac{t}{k}$. 
But we showed that $|S_i|<\frac{t}{k}$, contradiction.
Consequently
the initial assumption $|\overline{S_i}|> t$ is incorrect. 
End of proof of the claim. 


Let us assume (for a future contradiction) that there exists an algorithm 
${\cal A}_k$ that solve the $k$-set agreement problem in the system $\Sigma$ 
where (thanks to the claim) we have  $|\overline{S_i}|\leq t$ for any 
$i\in[1..(k+1)]$. \\

\vspace{-0.3cm}
\begin{figure}[htb]
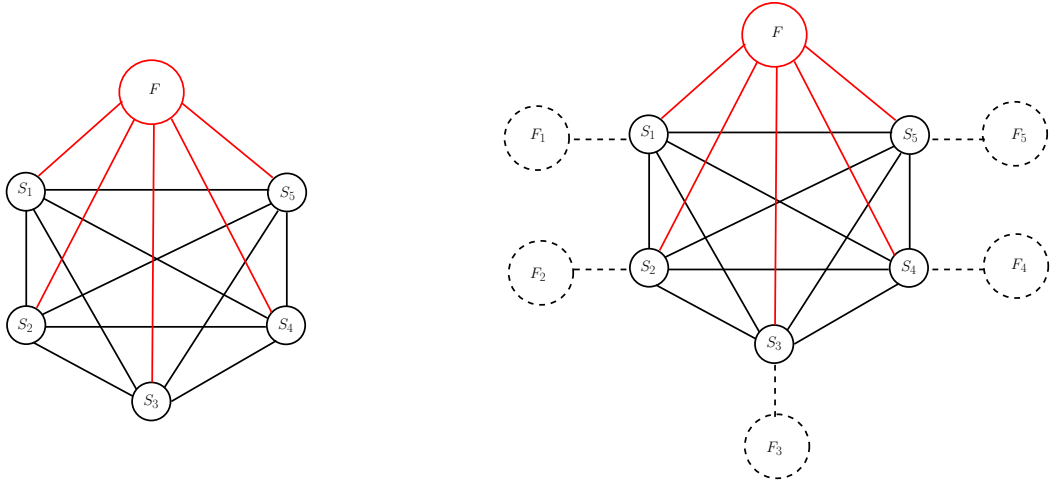

\begin{minipage}[l]{9cm}
\hspace{2cm}
\scalebox{0.23}{\input{fig-systeme-Sigma-1.pstex_t}}
\end{minipage}
\hspace{-0.5cm}
\begin{minipage}[r]{9cm}
\scalebox{0.23}{\input{fig-systeme-gamma.pstex_t}}

\end{minipage}
\caption{%
$\Sigma$ (left) and the behavior of the $t$ Byzantine processes  (right)}
\label{fig:system-sigma1-gamma}
\end{figure}

\noindent
Part 2 of the proof. \\
To specify the behavior of the Byzantine processes, 
let us consider  the right part of Figure~\ref{fig:system-sigma1-gamma}. 
The Byzantine processes of $F$ behave as follows. 
For each $i\in[1..(k+1)]$, the processes of $F$ 
simulate $(k+1)$ sets of processes, $F_1$, ..., $F_{k+1}$, such that
each set $F_i$ behaves correctly (i.e., execute ${\cal A}_k$) 
with respect to $S_i$.
We say that the processes of $F$ ``play $(k+1)$ duplicity roles''. 

Let us now suppose that the processes in $S_1\cup \cdots\cup  S_{k+1}$
execute  algorithm ${\cal A}_k$, while the processes of $F$ play the  
$(k+1)$ duplicity roles $F_1$, ..., $F_{k+1}$ described previously.
Moreover,  for each $i\in[1..(k+1)]$,
both the processes of $S_i$,  and the processes of $F$ 
in their $F_i$ role, propose the same value $v_i$,  these values 
being  such that $(i\neq j) \Rightarrow (v_i\neq v_j)$. 

As, for each $i$,  $|\overline{S_i}|\leq t$ (see the claim), 
it follows that, the processes of $S_i$ (which are correct) cannot distinguish 
the case where the processes of $F$ are Byzantine and play $(k+1)$ different 
roles while the processes of $\overline{S_i}$ are correct, from the case where 
the processes of $F$ are correct while the processes of $\overline{S_i}$ 
are Byzantine. Hence, as by assumption algorithm ${\cal A}_k$ 
is correct, it follows from its Termination and Validity properties that, 
for each  $i\in[1..(k+1)]$, the processes of $S_i$ decide $v_i$. Hence,
$(k+1)$ values are  decided by the  correct processes, which violates 
the Agreement property. Consequently, there is no algorithm ${\cal A}_k$.

While the previous reasoning  
relies on the fact that communication is by message-passing 
(Byzantine processes send different messages to each set $S_i$), it 
is independent of the fact that communication is synchronous or asynchronous. 
Hence, the proof is valid for both ${\cal BAMP}$ and ${\cal BSMP}_{n,t}$.
\renewcommand{\toto}{theo-necessity}
\end{proofT}

\vspace{-0.2cm}
\section{The Condition is Also Necessary in  ${\cal BARW}_{n,t}$}

\Xomit{
\paragraph{On ${\cal BARW}_{n,t}$ with respect  ${\cal BAMP}_{n,t}$}
Considering the model  ${\cal BAMP}_{n,t}$,  
let $p_x$ be  a correct process which broadcasts a message $m$, and 
$p_y$ a Byzantine process which claims that it never received $m$. 
There are executions in which this bad behavior of $p_y$ cannot be 
detected by  correct processes. 

The situation is different in the model  ${\cal BARW}_{n,t}$. Let us 
consider a correct process $p_x$ that writes a value $m$ in its SWMR register 
$\REG[x]$ and never overwrites it.  A Byzantine process $p_y$ cannot 
claim  that $p_x$ never wrote $m$. If it does, as any correct process 
can read $\REG[x]$,  $p_y$ can be discovered as being Byzantine by  
correct processes. 
It follows that, not to be discovered as being  faulty, a Byzantine process 
must report  values written by correct processes (be these reported values 
the original values written by the corresponding correct process, or 
falsified values). 
}

\begin{theorem}
\label{theo-necessity-arw}
There is no algorithm that solves $k$-set agreement
in ${\cal BARW}_{n,t}$ when $n\leq 2t+\frac{t}{k}$. 
\end{theorem}

\Xomit{
\paragraph{Remark}
Let us consider the model  ${\cal BAMP}_{n,t}$. 
While a correct process $p_x$ broadcast a message $m$, it is possible for 
a Byzantine process to claim forever that  it never received $m$. 
This behavior cannot be detected by a correct process. 

The situation is different in the model  ${\cal BARW}_{n,t}$. Let us 
consider a correct process $p_x$ that writes a value $m$ in its register 
$\REG[x]$ and never overwrites it.  A Byzantine process cannot 
claim forever that $p_x$ never wrote $m$. If it does, it will be
discovered as being Byzantine by the correct processes. 
It follows that, not to be discovered, a Byzantine process must not lie 
on the values written by correct processes.\\
}

\begin{proofT}
Considering the proof of Theorem~\ref{theo-necessity}, 
the proof consists in showing that the duplicity behavior of the Byzantine 
processes can be produced in ${\cal BARW}_{n,t}$. The theorem then follows
from the previous proof.  
%

Let $p_y\in F$, $\REG[y]$ a register that can be written only by $p_y$, and 
$p_x$ a correct process of a set $S_i$.  
The duplicity behavior of $p_y$  with respect to $p_x$ is produced as follows.
Just before $p_x$ reads $\REG[y]$, $p_y$ writes in  $\REG[y]$ the  
corresponding value produced by its execution of ${\cal A}_k$ in 
its $F_i$ role.  
It follows that, for each $i\in[1..(k+1)]$, the processes of  $F$ appear 
as correct processes to the processes of $S_i$, which concludes the proof. 
\renewcommand{\toto}{theo-necessity-arw}
\end{proofT}

\vspace{-0.2cm}
\section{The Byzantine Consensus Case $(k=1)$ in  ${\cal BSMP}_{n,t}$}
When considering consensus, the necessary condition  $n > 2t + \frac{t}{k}$
boils down to $n>3t$,  which has been shown to be both 
necessary and sufficient in the model ${\cal BSMP}_{n,t}$~\cite{LSP82}. 
It follows that, for $k=1$ and the model ${\cal BSMP}_{n,t}$, 
the proof of Theorem~\ref{theo-necessity} constitutes a new proof 
of the necessity of $n>3t$. A noteworthy feature of this  proof lies in 
the fact that it is a {\it direct} proof. Differently, 
the proofs of the condition $n>3t$ encountered in the literature 
(see the first proof given in~\cite{LSP82} or  classic  proofs 
given thereafter in  textbooks, e.g.,~\cite{L96,R10}) 
are decomposed in two steps: 
(a) first a proof showing that there is no consensus algorithm
in ${\cal BSMP}_{3,1}$,  
(b) followed by a simulation, on top of ${\cal BSMP}_{3,1}$, of an algorithm 
assumed to solve consensus in ${\cal BSMP}_{n,t}$ where $n\leq 3t$. 
%
In addition of being direct and based on  a classic {\it indistinguishability} 
argument,  the  proof of Theorem~\ref{theo-necessity} 
is more general as it considers a generic agreement problem, 
namely $k$-set agreement whose consensus is only its more constrained instance.

\section*{Acknowledgments}
\vspace{-0.2cm}
This work has been partially supported by the  
Franco-German DFG-ANR Project 40300781 DISCMAT
devoted to connections between mathematics and distributed computing, and 
the French  ANR project DISPLEXITY devoted to the study of computability 
and complexity  in distributed computing.


\end{document}